\documentclass[
 reprint,
 amsmath,amssymb,
 aps
]{revtex4-1}

\usepackage{graphicx}
\usepackage{dcolumn}
\usepackage{bm}
\usepackage{amsmath,amssymb}

\begin{document}

\title{Experimental study on the bifurcation of a density oscillator depending on density difference}

\author{Hiroaki Ito}%
\email{ito@chiba-u.jp}
\author{Taisuke Itasaka}
\author{Nana Takeda}
\author{Hiroyuki Kitahata}
\email{kitahata@chiba-u.jp}
\affiliation{Department of Physics, Graduate School of Science, Chiba University, Chiba 263-8522, Japan}%

\date{\today}

\begin{abstract}
Hydrodynamic instabilities often cause spatio-temporal pattern formations and transitions between them. We investigate a model experimental system, a density oscillator, where the bifurcation from a resting state to an oscillatory state is triggered by the increase in the density difference of the two fluids. Our results show that the oscillation amplitude increases from zero and the period decreases above a critical density difference. The detailed data close to the bifurcation point provide a critical exponent consistent with the supercritical Hopf bifurcation.
\end{abstract}

\maketitle

\section{Introduction}

Limit-cycle oscillations have been investigated intensively since they are related to a wide variety of dynamical phenomena both in natural and artificial systems. There are mainly two scenarios for the realization of oscillations. One is a harmonic-like oscillation realizing in an energy conservative system, and the other is a limit-cycle oscillation in which energy is alternately supplied and dissipated in one cycle. Thus, the observation of oscillation in open systems is the mark of an existence of a limit-cycle oscillation. Since energy dissipation is unavoidable in macroscopic and mesoscopic systems, many oscillatory phenomena observed in such systems can be regarded as limit-cycle oscillations.

One of the most famous experimental systems that exhibit limit-cycle oscillation is a chemical oscillation represented by Belousov-Zhabotinsky (BZ) reaction, in which oxidation and reduction processes occur alternately and the solution color changes according to such oxidation/reduction reactions~\cite{BZ,BZ2}. Other types of oscillators have also been reported; e.g., Briggs-Rauscher (BR) reaction, glycolysis, and electrolysis~\cite{Ertl}. Another class is hydrodynamical systems such as the B\'{e}nard convection system and the density oscillator. These hydrodynamical systems are often discussed in association with the atmospheric circulation and thermohaline circulation~\cite{Lappa,flow}. Martin firstly reported the density oscillator, and he discussed it as a simplified experimental system of thermohaline circulation~\cite{Martin}.

The limit-cycle oscillations in the hydrodynamical systems are important not only for the relation to such atmospheric and thermohaline circulations but also as the suitable model systems to investigate the bifurcation phenomena seen in dynamical systems. The hydrodynamical systems have many degrees of freedom, but they often show successive bifurcation structure from a trivial state to oscillatory states, quasi-periodic oscillation states, and chaotic states that can be described by a model with a small number of variables~\cite{convection_bif}. 

One of the simple and appropriate experimental systems that exhibit hydrodynamic limit-cycle oscillations is a density oscillator. 
In this system, the gravitational instability originating from an upset density profile of higher- and lower-density fluids induces upstream and downstream alternation, which can be regarded as a limit-cycle oscillation.
From Martin's first report in 1970, there have been experimental and theoretical papers on the density oscillators. Some are on the theoretical analysis on the mechanism of the oscillation~\cite{Steinbock,Okamura,Aoki,Kenfack,Ueno,BookKano,Kano,Kano2} and others are on the coupling between two-or-more oscillators~\cite{Yoshikawa,Horie,Gonza,Miyakawa1,Miyakawa2,Pikovsky}. For the bifurcation phenomenon, some theoretical studies have been performed based on the reduced ordinary differential equations, which have predicted various types of bifurcations according to various bifurcation parameters~\cite{Aoki,Kenfack} However, the experimental analyses on the bifurcation structure with physical parameters between the resting and oscillatory states are missing, and the detailed behaviors close to the bifurcation point remain unclear. 

In the present paper, we report the experimental results on the bifurcation between the resting and oscillatory states by changing the density of the higher-density fluid as a bifurcation parameter. We measured the oscillation amplitude and period depending on the density of the higher-density fluid, which indicate the bifurcation structure between the resting and oscillatory states. 

\section{Experimental setup}

\begin{figure}
    \centering
    \includegraphics{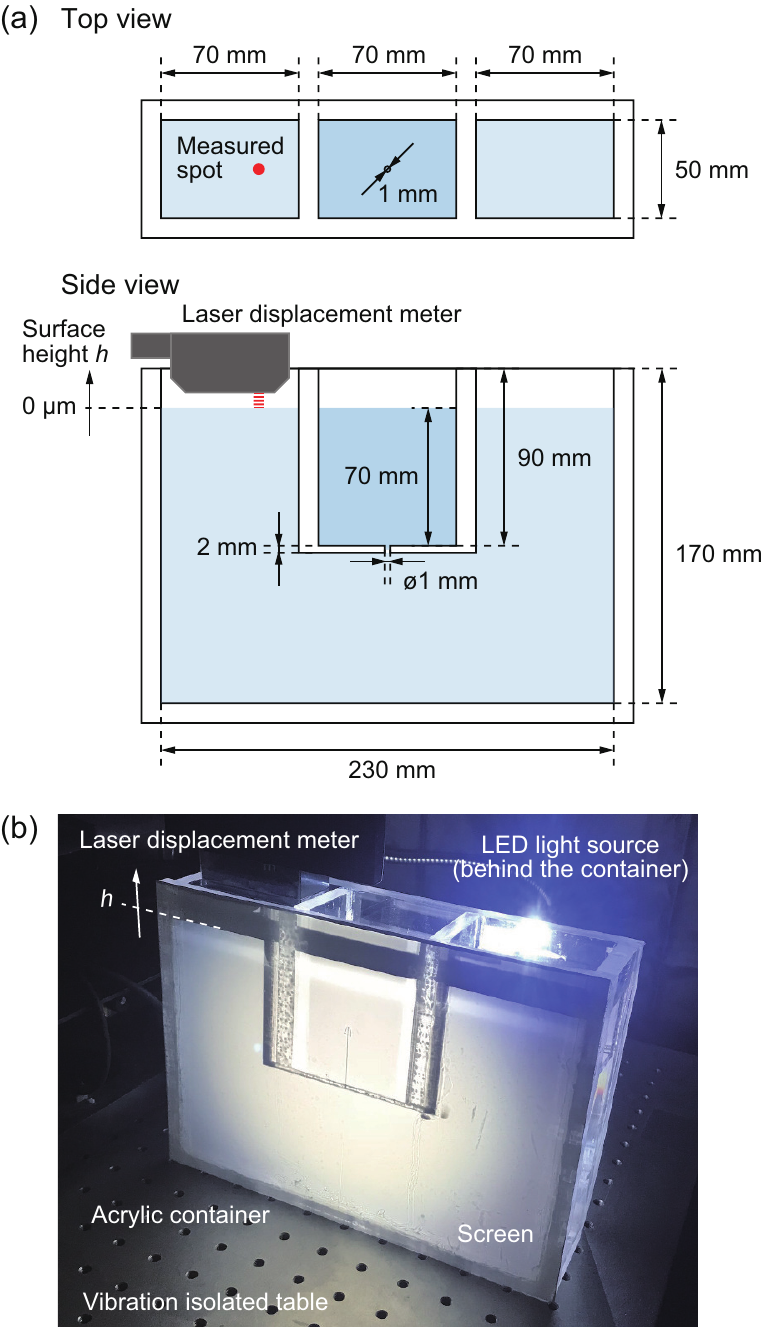}
    \caption{Experimental setup. (a) Dimensions of the chamber made of acrylic plates. The laser displacement meter and the measured spot are also shown. (b) Overview of the surface-height measurement and observation.}
    \label{fig:setup}
\end{figure}

All aqueous solutions and distilled water were prepared with Elix water purification system Elix UV 3 (Merck, Darmstadt, Germany). Sodium chloride (NaCl) aqueous solution as a higher-density fluid was prepared by dissolving NaCl (Wako Pure Chemicals, Tokyo, Japan) into the distilled water with various weight/volume concentrations $c$. All the aqueous solutions and pure water ($c = 0\,\mathrm{g/L}$, for control) were degassed for 1 hour in a vacuum chamber before use. 
An observation chamber of a density oscillator was made of acrylic plates, which is composed of a smaller container surrounded by a larger container for higher- and lower-concentration solutions, respectively. The acrylic plates were 10~mm in thickness except for the bottom of the smaller container with 2~mm thickness. The center of the bottom of the smaller container had a small cylindrical hole with a diameter of 1~mm and a length of 2~mm to connect these two containers. They were kept isolated by blocking the hole with a needle before the measurements. The dimensions of the observation chamber and a photograph of the experimental setup are shown in Fig.~\ref{fig:setup}.

We put 245~mL of NaCl aqueous solution in the smaller container, and 1400~mL of pure water in the larger container to reach the same water levels in these two containers. After removing the blocking of the small hole, the flow was induced through the hole if the concentration difference was greater than a critical value. The time series of the surface height of the solution in the outer container was measured with a laser displacement meter (32.77~fps, LT9010M, Keyence, Japan)~\cite{Ueno}.
We measured the surface height for 4000~s. As the surface height adequately relaxed to a steady resting or oscillatory state in 2000~s, we analyzed the data from 2000~s to 4000~s.
It should be noted that the steady oscillatory state with the repetitive macroscopic flow is characteristic for limit-cycle oscillations. To visualize the profile of the density difference between higher- and lower-NaCl concentration, we utilized the optical-index difference in a similar manner with the shadowgraph method~\cite{shadowgraph_do} using a light emitting diode (LED) as a light source and a plastic sheet (TAMIYA Inc., Shizuoka, Japan) as a screen. The obtained time series of the surface height were processed as follows: The surface height linearly decreased due to the evaporation, and thus the linear trend was reduced by subtracting the least-squares fitted linear function. Then, the data were smoothed by applying a band-pass filter between 0.003 and 0.019~Hz. Using the smoothed data and their time derivatives, we detected the local maximum and minimum points. The time series of the time derivative were calculated as the slope of adjacent $\pm10$ points (local data points for 0.64~s) obtained by the least-squares fitting. We obtained the mean peak-to-peak distances and the mean time intervals of the local minima as the oscillation amplitude and the period, respectively, for each condition of NaCl concentration $c$.

\section{Experimental results}

\begin{figure*}
    \centering
    \includegraphics{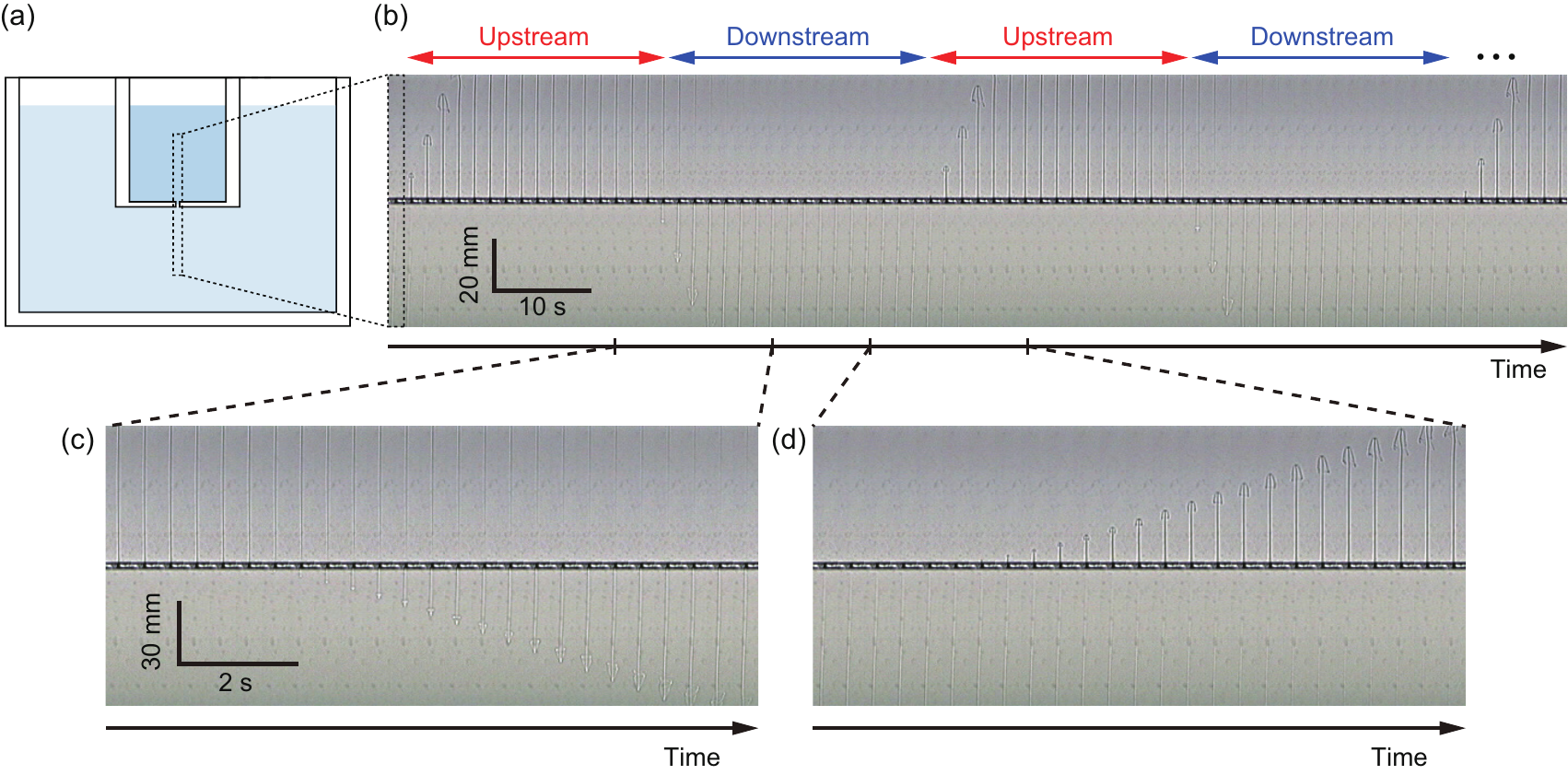}
    \caption{Snapshots of the oscillatory behavior of the density oscillator. The concentration of the NaCl aqueous solution was $c = 30\,\mathrm{g/L}$. 
    The oscillatory flow was visualized by the optical-index difference.
    (a) Displayed area. (b) Typical oscillatory flow. (c) Details for the switching from upstream to downstream. (d) Details for the switching from downstream to upstream. Snapshots every (b) 1.67~s and (c,d) 0.33~s are shown. The background-subtracted images are available in Fig.~S1 in the SM.}
    \label{fig:snapshot}
\end{figure*}

Figure~\ref{fig:snapshot} shows the upstream and downstream series through the small hole connecting the higher- and lower-concentration solutions. Characteristic oscillatory behavior of the density oscillator is observed. The concentration of the NaCl aqueous solution was $c = 30\,\mathrm{g/L}$. In this setup, the upstream of lower concentration solution and the downstream of higher concentration solution were visualized as dark and light images, respectively, since the cylindrical flow shape acts as concave or convex optical lenses. The upstream and downstream series were repeated with a typical period of $\sim 50~\mathrm{s}$ in this condition. In this timescale, the cylindrical flows did not mix with the surrounding solutions, i.e., diffusion of solutes is negligible. Figures~\ref{fig:snapshot}(b) and \ref{fig:snapshot}(c) show the details for the switching of these opposite flows, where the arrow-like shaped flow enters the other solution at an almost constant velocity.

\begin{figure}
    \centering
    \includegraphics{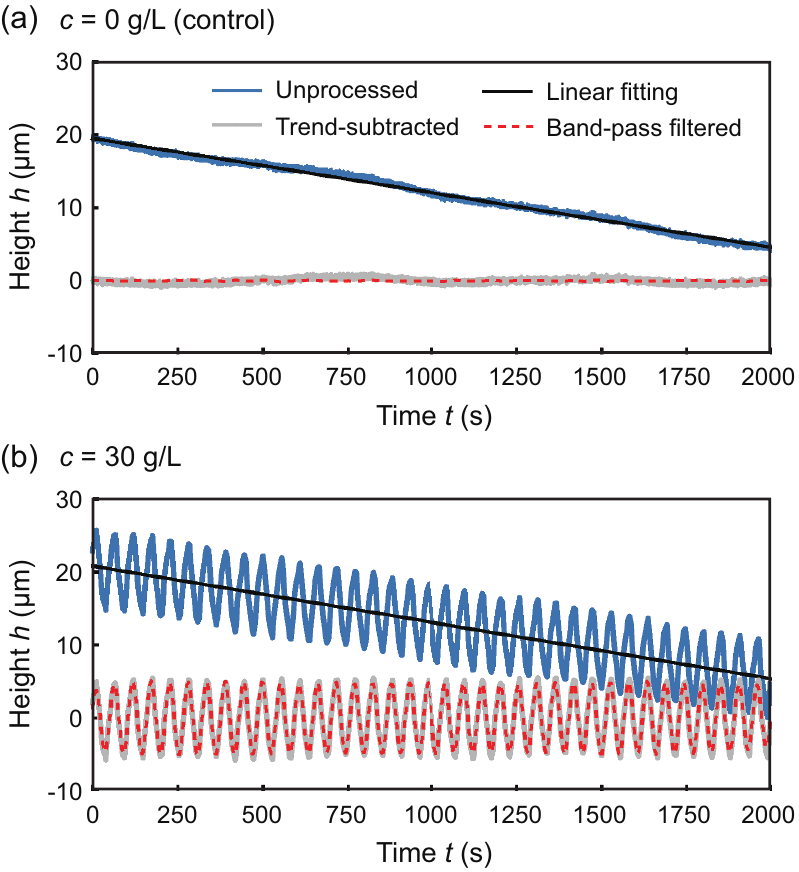}
    \caption{Time series of the surface height measured with a laser displacement meter for (a) $c = 0\,\mathrm{g/L}$ (control) and (b) $c = 30\,\mathrm{g/L}$. Unprocessed data (blue lines), linear fitting for the unprocessed data (black lines), trend-subtracted data (gray lines), and smoothed data with a band-pass filter (red-dashed lines) are shown.}
    \label{fig:tseries}
\end{figure}

For the quantitative measurement of the oscillation, we used a time series of the surface height of the solution in the outer container obtained by the laser displacement meter. The characteristic time series of the surface height is shown in Fig.~\ref{fig:tseries}. Figure~\ref{fig:tseries}(a) shows the result for $c = 0\,\mathrm{g/L}$ as a control condition. The unprocessed time series (blue line) shows a constantly-decreasing trend due to the slow evaporation of water, even though a total volume change of the aqueous phase is negligibly small. By subtracting the linear trend (black line), which is obtained by the least-squares method, time evolution of the surface height $h(t)$ is obtained. $h(t)$ directly reflects the fluid flow through the small hole between the inner and outer containers. While the time series for $c = 0\,\mathrm{g/L}$ is almost constant with small fluctuation (Fig.~\ref{fig:tseries}(a)), that for $c = 30\,\mathrm{g/L}$ exhibits steady oscillation (Fig.~\ref{fig:tseries}(b)), which corresponds to the oscillatory flow through the hole. Note that the linear trends for both resting ($c = 0\,\mathrm{g/L}$) and oscillatory ($c = 30\,\mathrm{g/L}$) states were measured in 3 hours under the condition with similar humidity, and showed the similar decreasing rate with the slopes of $-0.00746\,\mathrm{\mu m\cdot s^{-1}}$ and $-0.00772\,\mathrm{\mu m\cdot s^{-1}}$, respectively, which are compatible with the typical rate for water evaporation. For further quantification of the amplitude and period of the oscillation, we applied a band-pass filter in a frequency range 0.003--0.019~Hz (red-dashed line).

\begin{figure}
    \centering
    \includegraphics{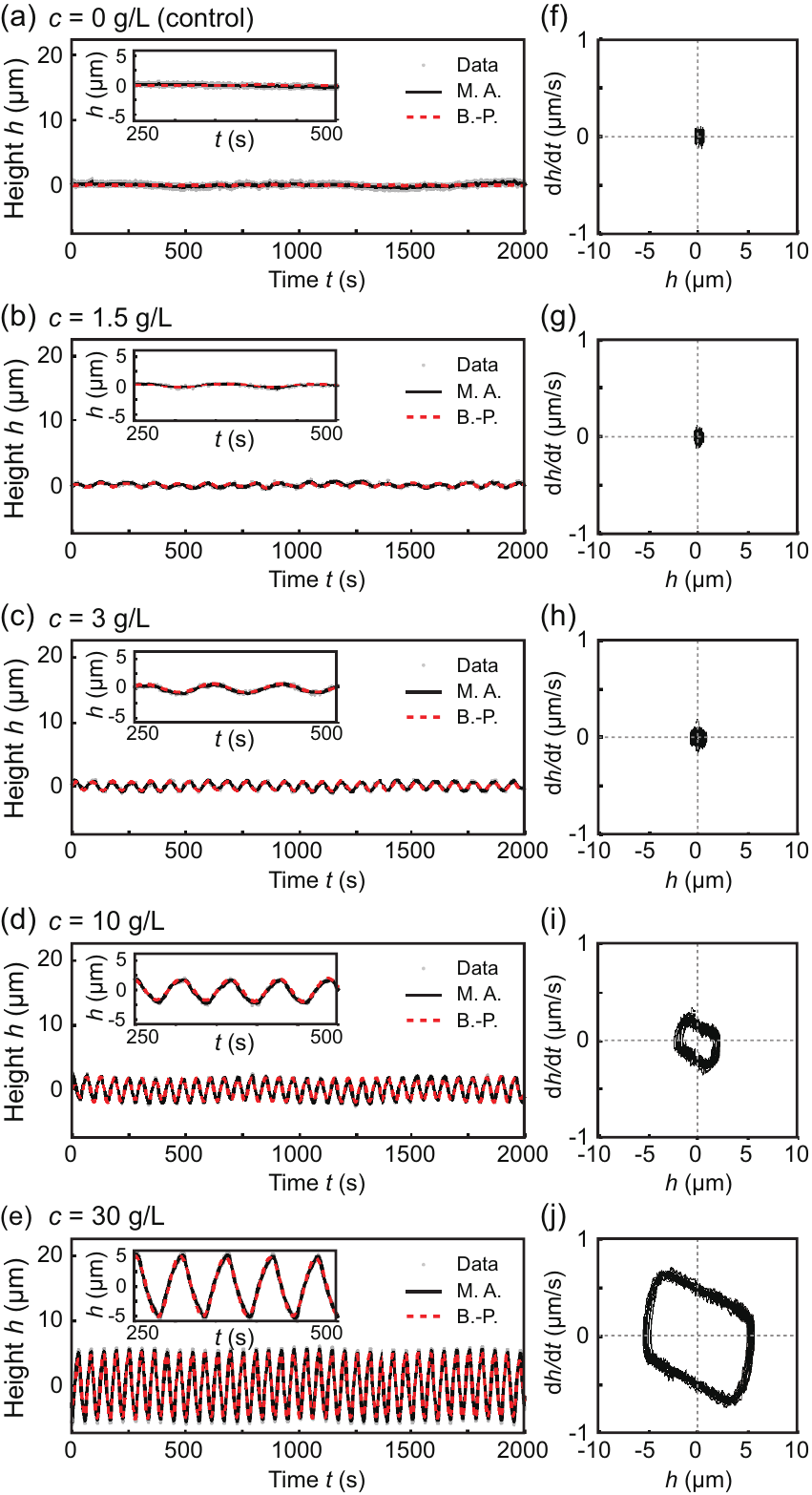}
    \caption{Oscillations for different NaCl concentrations $c$. (a) $0\,\mathrm{g/L}$, (b) $1.5\,\mathrm{g/L}$, (c) $3\,\mathrm{g/L}$, (d) $10\,\mathrm{g/L}$, and (e) $30\,\mathrm{g/L}$. Trend-subtracted data (Data, gray dots), moving-average (M. A., black lines), and band-pass filtered data (B.-P., red-dashed lines) are plotted. Each inset shows a magnified region of $250\,\mathrm{s}\leq t\leq 500\,\mathrm{s}$. (f--j) Corresponding trajectories in a phase space $(h,\mathrm{d}h/\mathrm{d}t)$.}
    \label{fig:conc_dependence}
\end{figure}

Figure~\ref{fig:conc_dependence} shows the concentration dependence of the typical time series of the surface height for $c = 0\,\mathrm{g/L}$ (control), $c = 1.5\,\mathrm{g/L}$, $c = 3\,\mathrm{g/L}$, $c = 10\,\mathrm{g/L}$, and $c = 30\,\mathrm{g/L}$. In addition to the original trend-subtracted data (gray points), we plotted smoothed data obtained by a moving average with adjacent $\pm64$ points (3.91~s) (black line) and the data with a band-pass filter (red-dashed line) (Figs.~\ref{fig:conc_dependence}(a-e)). The corresponding whole dynamics including the initial transient states are shown in Fig.~S2 in the SM. While the surface heights for small $c$, e.g., $c = 0\,\mathrm{g/L}$, are in a resting state, those for $c \geq 1.5\,\mathrm{g/L}$ exhibit characteristic oscillation. According to the increase in $c$, the amplitude becomes larger and the period becomes shorter. The detailed wave profiles, shown in each inset, also change from smooth sinusoidal-like oscillations to relaxation oscillations. Figures~\ref{fig:conc_dependence}(f--j) show the corresponding trajectories of the trend-subtracted smoothed data in a phase space $(h,\mathrm{d}h/\mathrm{d}t)$. According to the increase in $c$, the trajectories corresponding to the limit-cycle orbits with larger amplitudes appear. Also in the trajectories for $c \geq 10\,\mathrm{g/L}$, we can clearly see a characteristic shape of relaxation oscillation. Though the smoothed plot for small oscillation in $c = 1.5\,\mathrm{g/L}$ shows no clear cyclic trajectory due to the long-term fluctuation of the baseline, the plot after a band-pass filter shows sufficiently large oscillatory amplitude, which will be discussed later. 

\begin{figure}
    \centering
    \includegraphics{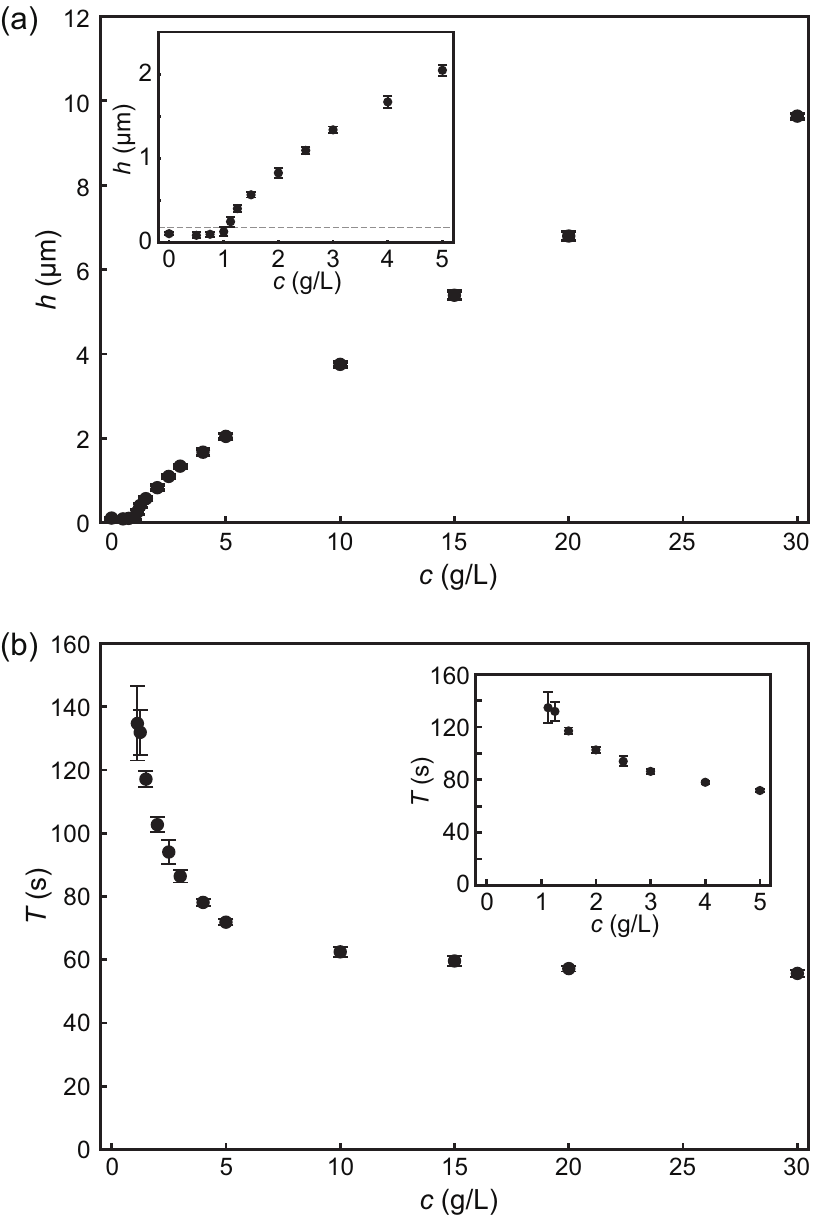}
    \caption{Amplitude and period of the density oscillator for various concentrations of NaCl aqueous solution as a higher-density fluid. (a) Amplitude. The dashed line in the inset shows the threshold value for the resting state. (b) Period. Each inset shows the magnified region around the bifurcation point.}
    \label{fig:amplitude_period}
\end{figure}

From the band-pass filtered data, we obtained the amplitudes and the periods of the oscillations for various concentrations $c$, as shown in Figs.~\ref{fig:amplitude_period}(a) and ~\ref{fig:amplitude_period}(b), respectively. The results in Fig.~\ref{fig:amplitude_period}(a) show that the amplitude shows a steep increase above a critical concentration $c_\mathrm{c}\simeq 1\,\mathrm{g/L}$, and it monotonically increases for larger concentrations. 
Here, we determined one standard deviation of the amplitude in the control condition ($c$ = 0~g/L) as the threshold amplitude between the resting and oscillatory states.
By this standard, $c \leq 1\,\mathrm{g/L}$ is in a resting state, and $c \geq 1.125\,\mathrm{g/L}$ shows a limit-cycle oscillation. For the conditions with limit-cycle oscillations $c \geq 1.125\,\mathrm{g/L}$, the oscillation period $T$ is plotted against various $c$ in Fig.~\ref{fig:amplitude_period}(b). The period $T$ monotonically decreases with the increase in the concentration $c$, and converges to $T \simeq 50\,\mathrm{s}$ for larger $c$.

\section{Discussion}
The driving force of the density oscillator is the gravitational energy. In fact, the density profile is upset in the initial stage; in other words, the higher-density fluid is located above the lower-density fluid. This instability can induce the limit-cycle oscillation through the dissipation of the gravitational energy.
From the experimental results, the limit-cycle oscillation was observed when the concentration of NaCl aqueous solution, $c$, was above a critical value $c_\textrm{c} \simeq 1 \, \textrm{g/L}$. Considering that the gravitational energy is proportional to the density, the results suggest that the oscillatory flow is induced by the upset density profile but it should suffer from the resistance due to the fluid viscosity. The gravitational energy due to the upset profile can overcome the viscosity for $c > c_\textrm{c}$, while it cannot for $c < c_\textrm{c}$. 

The existence of the critical value $c_\textrm{c}$ for the instability is also interpreted from the transport phenomenon. Since the solute transportation is dominated by diffusion in the resting state and by the advection in the oscillatory state, the Rayleigh number $Ra = (gL^3\Delta\rho/\rho)/(\nu D)$ provides a critical value for the density difference $\Delta\rho$ to drive the oscillation. Here, $g\,(\sim 10\,\mathrm{m\cdot s^{-2}})$ is the gravitational acceleration, $L\,(\sim 0.5\,\mathrm{mm})$ is the hole radius as a characteristic length, $\Delta\rho/\rho\,(\sim 10^{-2})$ is the ratio of the density difference, $\nu\,(\sim 1.0\times10^{-6}\,\mathrm{m^2\cdot s^{-1}})$ is the kinematic viscosity of water at room temperature, and $D\,(\sim 10^{-9}\,\mathrm{m^2\cdot s^{-1}})$ is the diffusion constant of solutes in water at room temperature. If we take a critical Rayleigh number $Ra\,(\sim 10^3)$ typical for Rayleigh-B\'{e}nard instability, the critical density difference $\Delta\rho_\textrm{c}$, namely, the critical solute concentration $c_\textrm{c}$ can be estimated as $c_\textrm{c} \sim 0.8\,\mathrm{g/L}$, which agrees well with the bifurcation point $\sim 1.0\,\mathrm{g/L}$ obtained in our experiment. Note that this value is much smaller compared to the parameter ranges in the previous experiments~\cite{Steinbock,Ueno}. Actually, the decreases in the density of salt water per a cycle are roughly estimated as $\sim -3\%$ in Steinbock et al.~\cite{Steinbock} and $\sim -0.2\%$ in Ueno et al.~\cite{Ueno}, which is much larger than our experiment, $\sim -0.003\%$. These previous experiments showed continuous shifts of the water surface in mm--cm length scale during the measurements due to the decrease in the density difference by mixing of the solutions between the two containers. In contrast, our experiment measured the motion of surface height in $\mathrm{\mu m}$ length scale, and therefore, the bifurcation phenomenon at the small density difference as well as the water evaporation could be detected.

\begin{figure}
    \centering
    \includegraphics{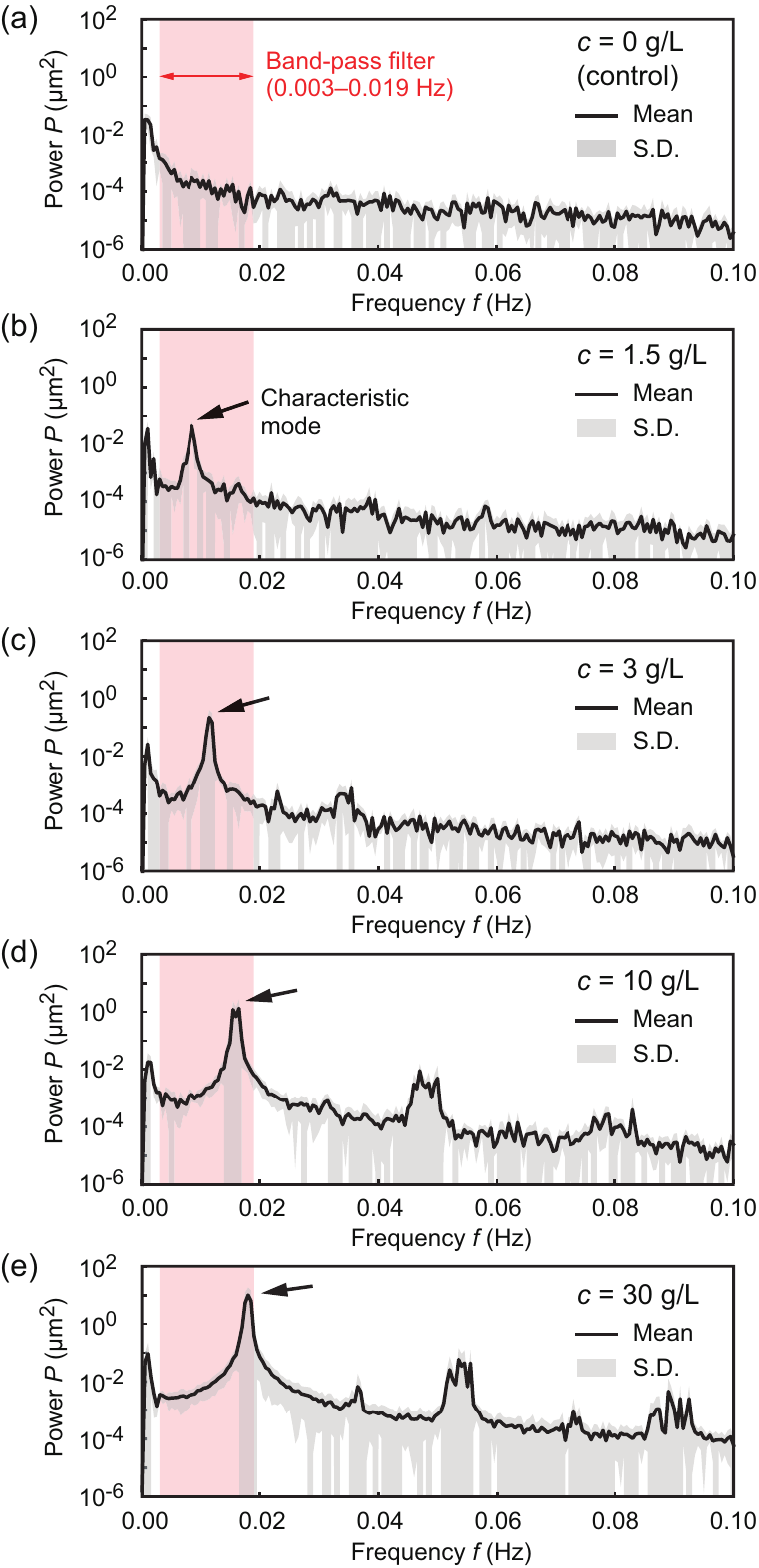}
    \caption{Power spectra for different NaCl concentrations $c$. (a) $0\,\mathrm{g/L}$, (b) $1.5\,\mathrm{g/L}$, (c) $3\,\mathrm{g/L}$, (d) $10\,\mathrm{g/L}$, and (e) $30\,\mathrm{g/L}$. Mean (black line) $\pm$ standard deviation (S.D., gray region) of the samples is plotted. Characteristic frequency modes for the limit-cycle oscillation are indicated by the black solid arrows. The spectra include the harmonics for these modes}
    \label{fig:power}
\end{figure}

\begin{figure}
    \centering
    \includegraphics{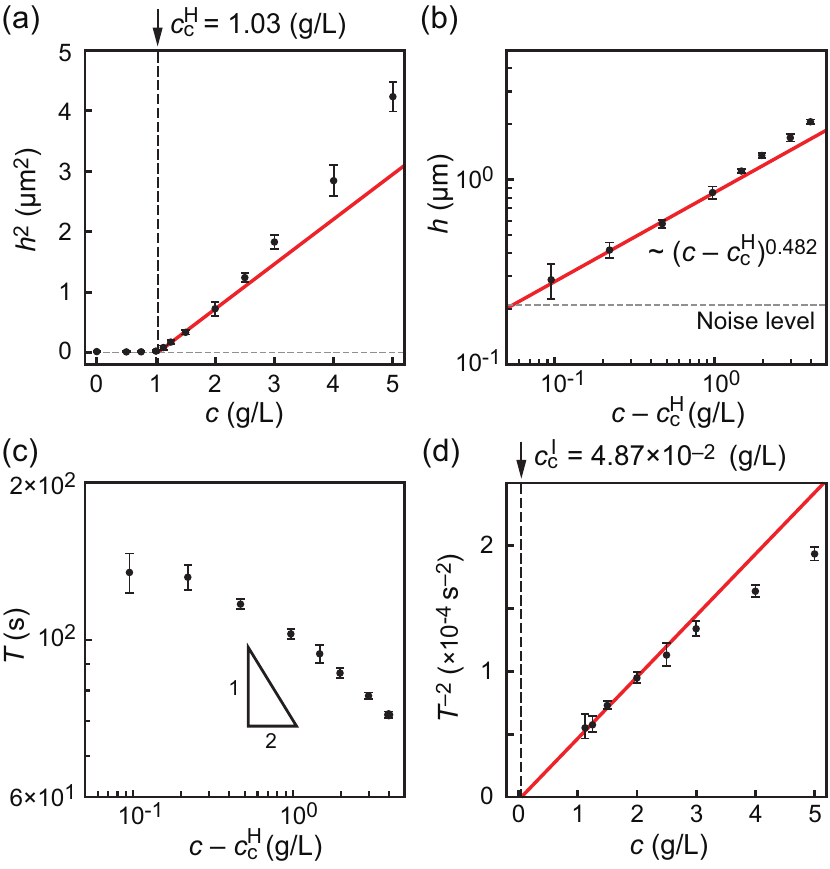}
    \caption{
    Critical behavior of the bifurcation with respect to the concentration. (a) Linear fitting of squared amplitude in $(1.25\,\mathrm{g/L}\leq c\leq 2\,\mathrm{g/L})$, close to the bifurcation point. The intersection estimates the Hopf bifurcation point $c_\mathrm{c}^\mathrm{H}=1.03\,\mathrm{g/L}$. (b) Double logarithmic plot of amplitude $h$ versus concentration difference $c-c_\mathrm{c}^\mathrm{H}$. The power exponent for $(1.25\,\mathrm{g/L}\leq c\leq 2\,\mathrm{g/L})$ is 0.482. (c) Double logarithmic plot of period $T$ versus concentration difference $c-c_\mathrm{c}^\mathrm{H}$. $T$ approaches constant as $c$ approaches the Hopf bifurcation point $c_\mathrm{c}^\mathrm{H}$. Slope is the eye guide for the comparison with the scaling $\sim (c-c_\mathrm{c}^\mathrm{H})^{-1/2}$. (d) Linear fitting of $T^{-2}$ in $(1.25\,\mathrm{g/L}\leq c\leq 2\,\mathrm{g/L})$ by assuming the scaling for the infinite-period bifurcation $\sim (c-c_\mathrm{c}^\mathrm{I})^{-1/2}$. The intersection estimates the possible infinite-period bifurcation point $c_\mathrm{c}^\mathrm{I} = 4.87\times10^{-2}\,\mathrm{g/L}$, which is apparently inconsistent with Fig.~\ref{fig:amplitude_period}(a).}
    \label{fig:exponent}
\end{figure}

From the viewpoint of the dynamical systems, this transition can be discussed in terms of bifurcation. There are several scenarios known for the transition from a resting state to an oscillatory state; supercritical Hopf bifurcation, saddle-node bifurcation of cycles (subcritical Hopf bifurcation), infinite-period bifurcation (saddle-node homoclinic bifurcation, saddle-node bifurcation on an invariant circle (SNIC)), homoclinic bifurcation, and so on \cite{Strogatz,Izhikevich,Guckenheimer,Kori}. 
For a supercritical Hopf bifurcation, the amplitude of oscillation raises from a bifurcation point in proportion to the square root of the distance from the bifurcation point, while the amplitude has a finite value at the bifurcation point for the other three types.

In the experiments, the amplitude of the oscillation seems to increase from zero with an increase in $c$ for $c > c_{\rm c}$ as shown in Fig.~\ref{fig:amplitude_period}(a). 
If we assume that the system undergoes the supercritical Hopf bifurcation, the dependence of the amplitude on the concentration $c$ is suggested to be proportional to $(c-c_\mathrm{c}^\mathrm{H})^{1/2}$ when the system is close to the bifurcation point. Here, $c_\mathrm{c}^\mathrm{H}$ is the critical concentration for the supercritical Hopf bifurcation.
We performed power-law fittings of the experimental data to determine the bifurcation point and the power exponent.
First, we checked the power spectra for each concentration $c$ to determine the range of the ``close-to-bifurcation-point region'', where the critical behavior is expected.
Figure~\ref{fig:power} shows the power spectra corresponding to the conditions shown in Fig.~\ref{fig:conc_dependence}.
Above $c_\mathrm{c}$, a clear first peak for the characteristic limit-cycle oscillation appears within a frequency range of the band-pass filter, 0.003--0.019~Hz.
For higher concentrations, $c\geq3\,\mathrm{g/L}$, harmonics originating from the nonlinear waveform of relaxation oscillation appear, i.e., the system is in the ``far-from-bifurcation-point region''.
Thus, we performed linear fitting of the squared amplitude in the ``close-to-bifurcation-point region'' as shown in Fig.~\ref{fig:exponent}(a), and obtained the critical concentration as $c_\mathrm{c}^\mathrm{H} = 1.03\,\mathrm{g/L}$.
Here, the closest oscillatory condition $c = 1.125\,\mathrm{g/L}$ was eliminated from the fitting, because the amplitude was close to the noise level. 
Then, the double logarithmic plot of the amplitude versus $c-c_\mathrm{c}^\mathrm{H}$ shown in Fig.~\ref{fig:exponent}(b) provides the power exponent close to the bifurcation point, resulting in $0.482$, which is close to $1/2$.
It suggests that the bifurcation would be classified into the supercritical Hopf bifurcation.

The supercritical Hopf bifurcation is also characterized by a finite angular velocity in the phase space at the bifurcation point. 
In Fig.~\ref{fig:amplitude_period}(b), the period increased as $c$ approached the bifurcation point from higher concentrations $c > c_\mathrm{c}$. 
This behavior recalls infinite-period bifurcation, in which the amplitude is finite while the period diverges at the bifurcation point $c_\mathrm{c}^\mathrm{I}$ with the scaling of $(c-c_\mathrm{c}^\mathrm{I})^{-1/2}$. We further checked the critical behavior of the period versus $c-c_\mathrm{c}^\mathrm{H}$ in the double logarithmic plot shown in Fig.~\ref{fig:exponent}(c), and confirmed that the period does not seem to diverge. Instead, it remains constant around the bifurcation point, which is also consistent with the supercritical Hopf bifurcation. Moreover, we also estimated another critical concentration $c_\mathrm{c}^\mathrm{I}$ by assuming the infinite-period bifurcation through the critical behavior of the inversed square period. This test resulted in $c_\mathrm{c}^\mathrm{I} = 4.87\times10^{-2}\,\mathrm{g/L}$ as shown in Fig.~\ref{fig:exponent}(d), which is apparently inconsistent with the experimentally indicated $c_\mathrm{c}\simeq1.0\,\mathrm{g/L}$ as observed in the amplitude shown in Fig.~\ref{fig:amplitude_period}(a). While our experimental results and analyses strongly support the supercritical Hopf bifurcation, the precise asymptotic behaviors are still not clear due to the experimental limitations. 
From the experimental observation only, it is difficult to completely exclude the possibilities for other types of bifurcations.
From the present results, the saddle-node bifurcation of cycles was dismissed since the bistability between the resting and oscillatory states was not observed.
Further study is needed to identify the bifurcation class of the density oscillator depending on the density difference.

\section{Summary}
We investigated the transition from the resting state to the oscillatory state depending on the density difference in a density oscillator. The limit-cycle oscillation, where the upstream and downstream alternations occur, was observed for the higher density difference, while no hydrodynamic flow was observed for the lower density difference. 
The detailed data close to the bifurcation point provide a critical exponent close to $1/2$ and a finite period around the bifurcation point, which is consistent with the supercritical Hopf bifurcation.
Further experimental and theoretical studies should be performed to exactly identify the bifurcation class since the experimental results still showed the ambiguity.

The density oscillator is an excellent experimental system for the limit-cycle oscillation with hydrodynamic instability since it can be treated from the standpoint of physics; complex processes such as chemical reaction and phase transition are not included, and thus only hydrodynamics is taken into consideration.
The present experimental study will give fundamental knowledge on the nonlinear oscillation with hydrodynamic instability from the viewpoint of bifurcation theory in dynamical systems, and help further studies using density oscillators.

\begin{acknowledgments}
This work was supported by JSPS KAKENHI Grant Numbers JP19K14675, JP16H03949. It was also supported by the Japan-Poland Research Cooperative Program ``Spatio-temporal patterns of elements driven by self-generated, geometrically constrained flows'' and the Cooperative Research Program of ``Network Joint Research Center for Materials and Devices'' (No.~20191030).
\end{acknowledgments}

\newpage
\renewcommand{\thefigure}{S\arabic{figure}}
\setcounter{figure}{0}
\begin{figure*}[b]
    \begin{minipage}{\textwidth}
    \centering
    \includegraphics{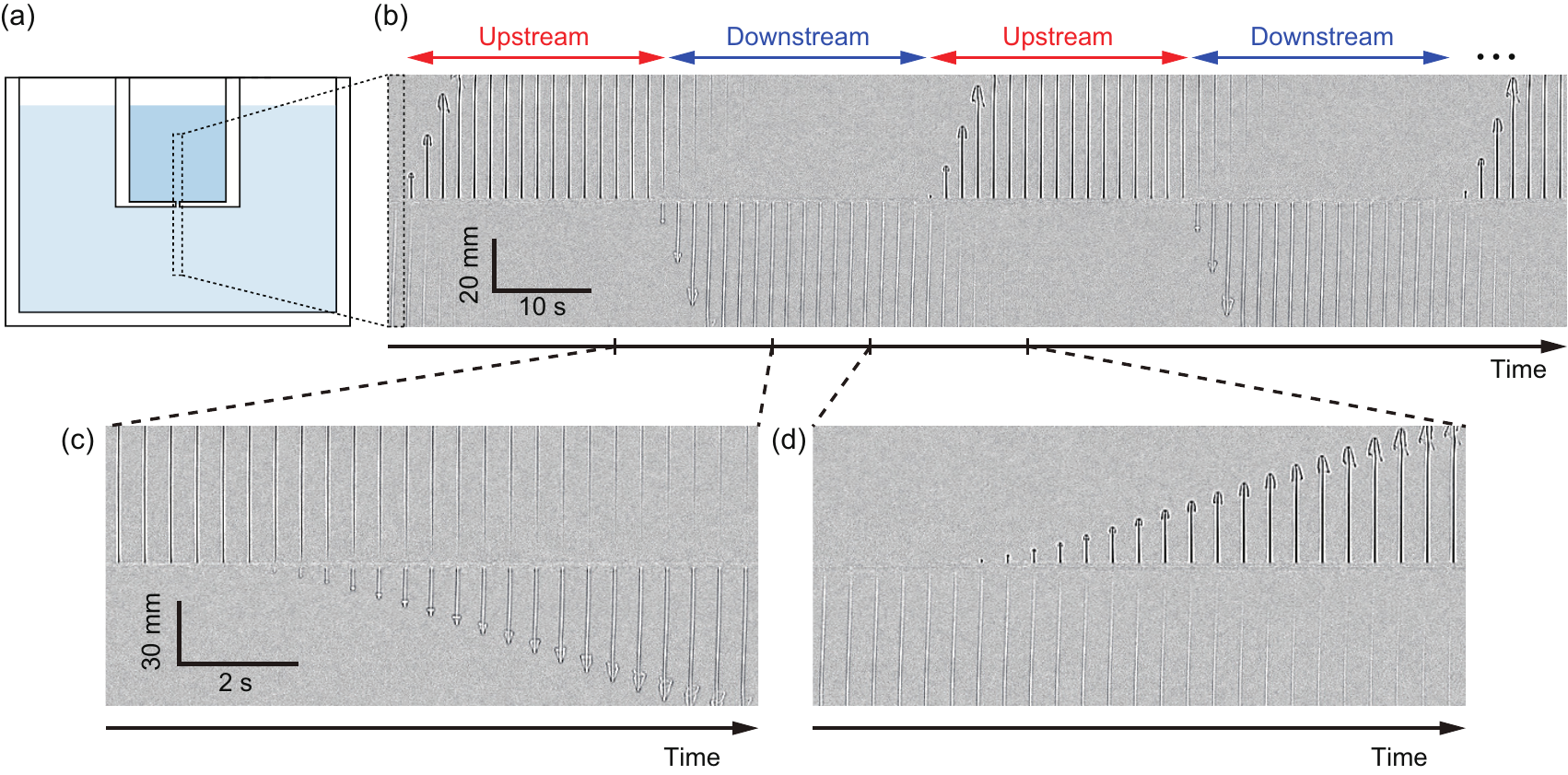}
    \caption{Background-subtracted images corresponding to Fig.~2 in the main text. The concentration of the NaCl aqueous solution was $c = 30\,\mathrm{g/L}$. (a) Displayed area. (b) Typical oscillatory flow. (c) Details for the switching from upstream to downstream. (d) Details for the switching from downstream to upstream.}
    \label{fig:s1}
    \end{minipage}
\end{figure*}

\begin{figure*}[b]
    \centering
    \includegraphics{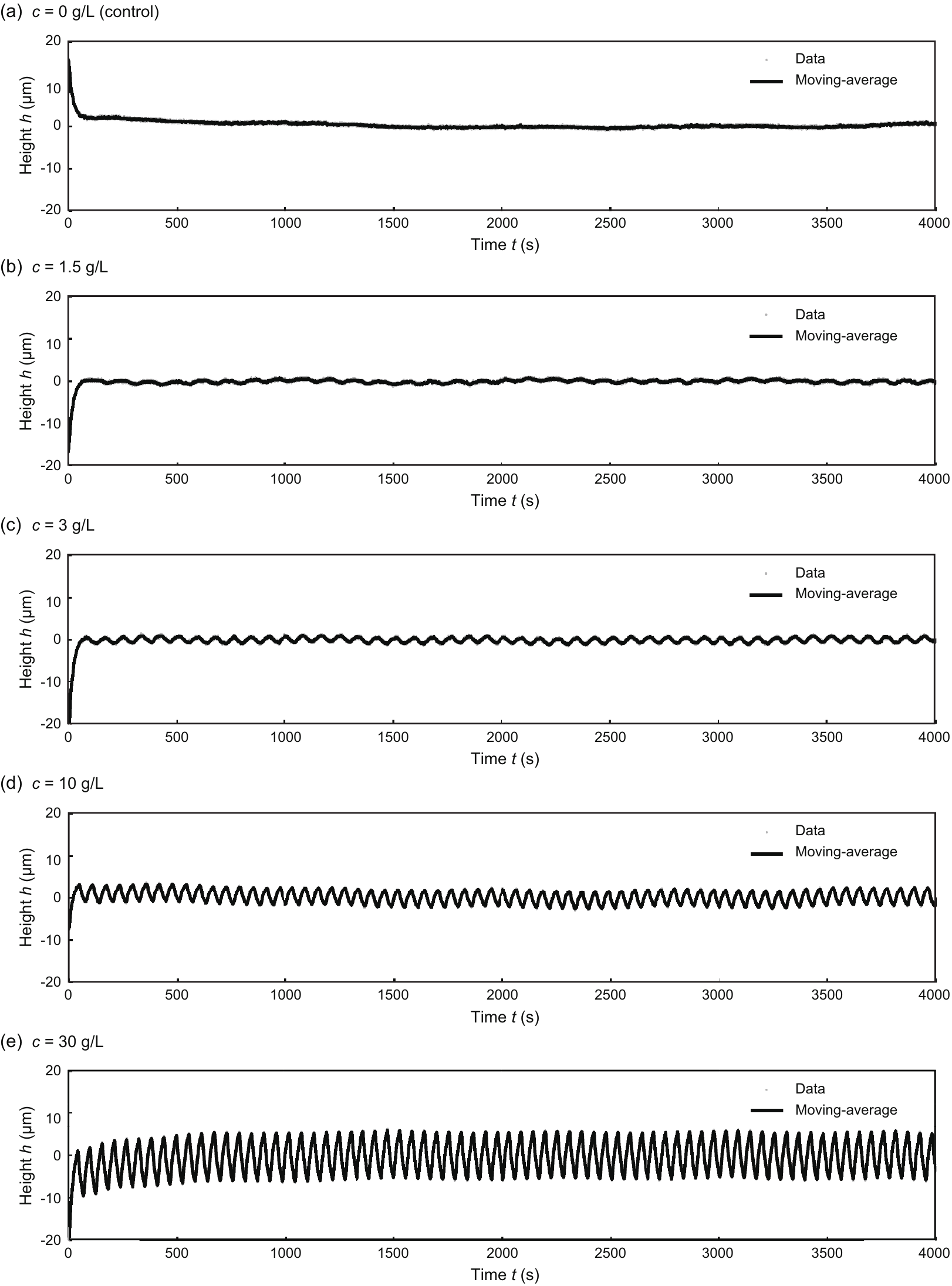}
    \caption{Typical results of the surface-height measurements for different NaCl concentrations $c$. (a) $0\,\mathrm{g/L}$, (b) $1.5\,\mathrm{g/L}$, (c) $3\,\mathrm{g/L}$, (d) $10\,\mathrm{g/L}$, and (e) $30\,\mathrm{g/L}$. Trend-subtracted data (Data, gray dots) and moving-averages (black lines) are plotted. The system reached the steady states within $\sim1000\,\mathrm{s}$. Each trend was obtained by the linear fitting for $t = 2000$--$4000\,\mathrm{s}$.}
    \label{fig:s2}
\end{figure*}

\end{document}